\documentclass[12pt]{article}
\usepackage{epsf}
\usepackage{rotate}
\def\lsim{\mathrel{\rlap {\raise.5ex\hbox{$ < $}}
{\lower.5ex\hbox{$\sim$}}}}
\def\gsim{\mathrel{\rlap {\raise.5ex\hbox{$ > $}}
{\lower.5ex\hbox{$\sim$}}}}
\def\np#1#2#3{{\it {Nucl. Phys.}} {\bf{B#1}} (#2) #3}
\def\pl#1#2#3{{\it {Phys. Lett.}} {\bf{B#1}} (#2) #3}
\def\prl#1#2#3{{\it {Phys. Rev. Lett. }}{\bf{#1}} (#2) #3}
\def\pr#1#2#3{{\it {Phys. Rev.}} {\bf{D#1}} (#2) #3}
\def\prep#1#2#3{{\it {Phys. Rep.}} {\bf{#1}} (#2) #3}
\def\nl{\hfil\break}
\begin{document}
\begin{titlepage}
\begin{flushright}
SISSA/173/96/EP\\
hep-ph/9702295{\hskip.5cm}\\  
\end{flushright}
\begin{centering}
\vspace{.3in}
{\bf STRING SCALE UNIFICATION IN AN 
$SU(6){\times}SU(2)$ GUT }\\
\vspace{2 cm}
{J. RIZOS$^{1,2}$
 and K. TAMVAKIS$^2$}\\
\vskip 1cm
{$^1 $\it International School for Advanced Studies, SISSA}\\
{\it Via Beirut 2-4, 34013 Trieste, Italy}\\
\vskip 1cm
{$^2$\it {Physics Department, University of Ioannina\\
Ioannina, GR45110, GREECE}}\\

\vspace{1.5cm}
{\bf Abstract}\\
\end{centering}
\vspace{.1in}
 We construct and analyze an $SU(6)\times SU(2)$ GUT. 
The model is k=1 string embedable in the sense that
we employ only chiral representations allowed at the 
k=1 level of the associated Ka\v{c}-Moody Algebra. 
Both cases $SU(6)\times SU(2)_{L}$ and $SU(6)\times SU(2)_{R}$ 
are realized. The model is characterized by the 
$SU(6)\times SU(2)\rightarrow SU(4)\times SU(2)\times SU(2)$ 
breaking scale $M_X$, and the 
$SU(4)\times SU(2)\times SU(2)\rightarrow 
SU(3)_{C}\times SU(2)_{L}\times U(1)_{Y}$ 
breaking scale $M_{R}$ . The spectrum bellow $M_R$ includes 
an extra pair of charge-1/3 colour-triplets of 
mass $M_{I}\leq M_{R}$ that does not couple to matter fields
 and, possibly, an extra pair of isodoublets.
 Above $M_{X}$ the $SU(6)$ and $SU(2)$ gauge couplings 
always unify at a scale which can be taken to be the string 
unification scale $M_{s}\sim 5\times 10^{17} GeV $.
 The model has Yukawa coupling unification  since
quarks and leptons obtain their masses 
from a single Yukawa coupling. Neutrinos obtain acceptably 
small masses through a see-saw mechanism.
 Coloured triplets that couple to matter fields are naturally 
split from the coexisting isodoublets without the need of 
any numerical fine tuning. 
\vspace{1cm}
\begin{flushleft} 
SISSA/173/96/EP\\
February 1997\\
\end{flushleft}
\hrule width 6.7cm \vskip.1mm{\small \small}
 \end{titlepage}
\begin{bf}1.Introduction.\end{bf}

 Superstring Theory \cite{GSW} is at present our best 
candidate for a unified theory of all particle 
interactions being a consistent theoretical 
framework that incorporates quantum gravity and 
supersymmetric gauge theories \cite{NHK}. Unification takes place
at energies close to the Planck scale while at 
lower energies gauge interactions are described in 
terms of an effective field theory whose spectrum
consists of the massless string modes. 
 From the low energy point of view the Standard Model and its
$N=1$ supersymmetric extension, the Minimal Supersymmetric Standard Model 
(MSSM),  can be naturally embedded in a Grand Unified Theory (GUT)
with interesting phenomenological and cosmological consequences.
GUTs \cite{GUTS} can successfully predict $\sin^2\theta_W$, 
fermion mass relations,
 charge quantization
as well as provide a mechanism to explain 
the baryon asymmetry of the Universe \cite{BA}.
However, in the context of quantum field theory no severe restrictions exist 
on the gauge group or the matter content of a GUT apart from the requirement
 that it should accommodate the MSSM. Thus a lot of possibilities arise 
including minimal SU(5) and its extensions \cite{SU5}, varieties 
of SO(10) \cite{SO10}, E(6) models \cite{E6} e.t.c. 

 Accommodating a GUT in the framework of Superstring Theory
\footnote{The term GUT is used here in a loose sense for a gauge model with
gauge group $G\supset SU(3)\times SU(2) \times U(1)$ which accomodates
the MSSM particles and displays partial
unification and/or correlation  of the standard model group gauge factors.}, or more precisely
assuming that the GUT is the low energy effective field
theory of a four dimensional heterotic superstring compactification, 
imposes serious 
restrictions on the GUT spectrum. For gauge groups realized at level $k=1$
of the  World-Sheet Affine Algebra, only the chiral multiplets in the 
vector and antisymmetric representations
of $SU(n)$ groups and the vector and spinor of $SO(n)$ groups are massless.
The absence of adjoint scalars limits the possibilities of breaking to
the MSSM through the Higgs mechanism and diminishes the number of candidate
GUT models. Apart from this serious restriction, superstrings  offer
a new possibility. The 
GUT gauge group does not have to be simple in order to
guarantee unification. Semi-simple or product groups are equally  
acceptable since string theory takes over the job of gauge coupling unification.
 
Historically the first GUTs that were studied as candidate low energy superstring
effective models were the maximal $E(6)$ subgroups \cite{SIE6} as they arise 
naturally in the 
context of Calabi-Yau compactifications \cite{KY1,KY3,RV}. A typical example of 
such a   
model is the $SU(3)\times SU(3)\times SU(3)$ GUT \cite{SU33} while other 
possibilities as $SU(6)\times U(1)$ have also been considered \cite{PANA}. Soon 
it was realized that $SO(10)$ subgroups were also possible in the context
of Orbifolds \cite{ORBI,RV} or in the 
{\it{free-fermionic formulation}} \cite{FFF}. Two
typical GUT examples, namely the ``flipped- SU(5)" \cite{BN} and the 
supersymmetric
Pati-Salam model \cite{AL}, were explicitly constructed in the framework of
the Fermionic Superstrings \cite{NA,ALR}. Standard Model-like alternatives 
have also been explored \cite{FAR}. 
 
Recently, a new the possibility has been explored 
, namely GUTs realized at 
the $k\ge2$ level of 
heterotic superstring compactifications \cite{BDS}. Some
generic features of these models have been presented in \cite{IBA,DIENES}. 
Higher level
models allow the presence of adjoint chiral superfields and
 thus the number of the candidate GUT gauge groups is enhanced.
Although  some explicit superstring  models
have been presented  \cite{MKT},  there are still some problems to
overcome \cite{IBA,DIENES,KOSTAS}.
 Such is the absence of certain couplings. For instance,
the remnants of the GUT breaking adjoint Higgs, an isotriplet and 
a colour-octet, remain light to all orders in perturbation
theory \cite{IBA,KOSTAS}. 
 
Another issue raised in the framework of the Superstring embedding of the MSSM
is that of the ``unification mismatch". Assuming MSSM to hold, low energy data 
indicate gauge coupling unification at a scale $M_{X}\sim 10^{16} GeV$ \cite{USSM}
 more 
than one order of magnitude below the string scale $M_{s}\sim 5\times 10^{17} GeV$
which is the typical string unification scale. The prospects to fill this gap
by threshold corrections  due to  the infinite tower of massive string modes
 \cite{THRE1,THRE2} seem rather restricted \cite{ILR} as explicit calculations
in superstring models show that such thresholds are  very small \cite{STHRE}.
 This could be taken as an
indication for the existence of {\em intermediate scales} in which extra matter
states become massive and their thresholds increase the coupling unification 
scale
up to $M_s$ \cite{ES1,ES2,KOSTAS}. One is thus motivated to incorporate this 
additional feature
in a candidate GUT model.

In the current state of elaboration of String Theory, the string vacuum is
not uniquely determined and the various gauge groups appear on
equal footing. It is hence interesting, from the low energy point of view,
 to classify all gauge models
satisfying the known string constraints and see whether they can
meet the criteria imposed by low energy data and at the same time
look for signatures that distinguish among them.

 As is well known, all quark and lepton fields can be 
accommodated in the ${\bf{27}}$ representation of $E(6)$ 
together with a pair of isodoublets suitable to serve as 
electroweak Higgses and a pair of charge $\pm {\frac {1}{3}}$ 
colour-triplets. Among the maximal subgroups of $E(6)$ we find 
$SU(6)\times SU(2)$, 
a regular subgroup which apart from \cite{SABAS} has not received much 
attention in the literature.
Decomposing the ${\bf 27}$ under 
$SU(6)\times SU(2)$ we get \cite{SL}
$${\bf 27}=({\bf {\overline{6}}},{\bf 2})+({\bf 15},{\bf 1})$$
where ${\bf 15}$ is the two-index antisymmetric representation 
of $SU(6)$. In what follows we shall construct a $k=1$ string 
embedable $SU(6)\times SU(2)$ GUT using chiral superfields 
in the $({\bf 6},{\bf 2})$ and $({\bf 15},{\bf 1})$ 
representations only. 
Two distinct ways of embedding the standard model group in 
$SU(6)\times SU(2)$ arise depending on whether we identify the 
electroweak $SU(2)_{L}$ with the GUT $SU(2)$ factor or with 
the $SU(4)\times SU(2)$ subgroup of $SU(6)$. In the latter 
case we have an $SU(6)\times SU(2)_{R}$ model while in the 
former an $SU(6)\times SU(2)_{L}$ model. 
Both cases $SU(6)\times SU(2)_{L}$ 
and $SU(6)\times SU(2)_{R}$ are realizable. Apart from
 the GUT scale $M_{X}$ 
of $SU(6)$ breaking, the model is also characterized by 
the extra scale of $SU(2)_{R}$ breaking $M_{R}$. 
Below the scale $M_{R}$ the spectrum contains an extra 
pair of coloured triplets of mass $M_{I}\leq M_{R}$.
 The $SU(6)$ and $SU(2)$ gauge couplings always meet at 
energies above $M_{X}$ introducing an additional scale 
which can always be taken to be the string unification 
scale $M_{s}\sim 5\times 10^{17} GeV $. 
In addition these models have some other 
 very interesting features. 
The electroweak Higgses, together with extra charge-1/3 
colored triplets, participate in the same representations 
that contain quarks and leptons. These isodoublets 
and colour-triplets also come in family replicas.
 Despite that, the triplet-doublet splitting can be 
achieved naturally without the need of any numerical 
fine tuning. Quark and lepton masses arise from a 
single Yukawa coupling, a cubic interaction of matter 
superfields, just as in $E_{6}$, despite 
the fact that matter fields come in a reducible 
representation. Thus, $SU(6)\times SU(2)$ displays 
Yukawa coupling unification. In the case of one pair of 
Higgs isodoublets the 
mass relations $m_{t}cot{\beta}=m_b=m_{\tau}$ are true and 
imply strong constraints 
on $m_t$, $tan{\beta}$ \cite{LLL} and the various supersymmetry breaking 
parameters \cite{WWW}. 

\bigskip
 {\begin{bf}2. Symmetry breaking and model building.\end{bf}}

 Decomposing the ${\bf 15}$ representation  
under  $SU(4)\times SU(2)\times U(1)$ $\subseteq$ 
$SU(6)$ we obtain 
$${\bf 15}=({\bf 1},{\bf 1},4)+({\bf 6},{\bf 1},-2)+
({\bf 4},{\bf 2},1)$$
Thus, it is clear that the GUT symmetry breaking 
$SU(6)\times SU(2)\rightarrow SU(4)\times SU(2)
\times SU(2)$ can be achieved with a non-zero v.e.v. of 
$({\bf 15},{\bf 1})+({\bf {\overline{15}}},{\bf 1})$ in 
the $D$-flat direction $\langle ({\bf 1},{\bf 1},4)\rangle 
=\langle ({\bf 1},{\bf 1},-4)\rangle $ . The decomposition 
of the adjoint ${\bf 35}$ of $SU(6)$ under $SU(4)\times 
SU(2)\times U(1)$ takes the form ${\bf 35}=
({\bf 1},{\bf 1},0)+({\bf 1},{\bf 3},0)
+({\bf 15},{\bf 1},0)+({\bf 4},{\bf 2},-3)+
({\bf {\overline{4}}},{\bf 2},3)$. This 
helps to see that the tetraplets in ${\bf 15}$ 
 will be absorbed while  the sextets $({\bf 6},{\bf 1},-2)$ , 
$({\bf {6}},{\bf 1},2)$ and the singlet 
combination $({\bf 1},{\bf 1},4)+({\bf 1},{\bf 1},-4)$ 
will survive the Higgs phenomena. Nevertheless, cubic superpotential 
couplings 
$$({\bf 15})^{3}=
({\bf 6},{\bf 1},-2)^{2}({\bf 1},{\bf 1},4)+
({\bf 4},{\bf 2},1)^{2}({\bf 6},{\bf 1},-2)$$
while not spoiling $F$-flatness, can give the sextets a mass. 
The first stage of symmetry 
breaking being as described, two possibilities arise as we 
move further down. 

\medskip
\begin{it}a) An $SU(6)\times SU(2)_{L}$ model.\end{it}

 The desired breaking pattern down to the standard 
model gauge group is 
 $$SU(6)\times SU(2)_{L}{{M_X}\atop\longrightarrow}
SU(4)\times SU(2)_{R}\times SU(2)_{L}{{M_R}\atop\longrightarrow}
SU(3)_{C}\times U(1)_{Y}\times SU(2)_{L} $$ 
 As we described above, the first stage of symmetry breaking
requires a pair of Higgses 
\footnote{We introduce here a mixed notation where the 
 $SU(4)\times SU(2)_{R}\times SU(2)_{L}\subset SU(6)\times SU(2)$
quantum numbers are shown explicitly
while the symbols for the fields  indicate 
Standard Model quantum numbers in the usual notation : $\Delta^c, D^c,
\delta^c\to({\bf\bar3},{\bf1},\frac{1}{3})$, $U^c,u^c\to
({\bf\bar3},{\bf1},-\frac{2}{3})$, $E^c,e^c\to({\bf1},{\bf1},1)$,
$ N,n,\nu^c\to({\bf1},{\bf1},0)$,   
$h,L,\eta\to({\bf1},{\bf2},-\frac{1}{2})$ etc.  }
\begin{eqnarray}
{\cal H}({\bf {\overline{15}}},{\bf 1})+
{\overline{\cal H}}({\bf 15},{\bf 1})&=&
{(N_{H})}_{({\bf 1},{\bf 1},{\bf 1})}+
{({\Delta^{c}_{H}}+{\Delta_{H}})}_{({\bf {\overline{6}}},{\bf 1},{\bf1})}
\nonumber\\
&&\mbox{}+({E^{c}_{H}+N^{c}_{H}+D^{c}_{H}
+U^{c}_{H})}_{({\bf {\overline 4}},{\bf 2},{\bf1})} +conj. reps.
\end{eqnarray}
 A $D$-flat v.e.v. $\langle N_{H} \rangle =
\langle {\overline{N}}_{H}\rangle=V_{X}$ breaks 
$SU(6)\times SU(2)_{L}$ to 
$SU(4)\times SU(2)_{R}\times SU(2)_{L}$. The surviving 
states are the sextets ${(\Delta^{c}_{H}+
\Delta_{H})}_{({\bf 6},{\bf 1},{\bf1})}$, 
${({\overline{\Delta}}^{c}_{H}+
\overline{\Delta}_{H})}_{({\bf 6},{\bf 1},{\bf1})}$ plus a singlet 
combination $N_{H} + \overline{N}_{H}$.

 The second stage of symmetry breaking down to the standard model 
gauge group can be achieved introducing an additional pair 
\begin{eqnarray} 
H({\bf {\overline{15}}},{\bf 1}) +\overline H
({\bf 15},{\bf 1})&=&{(n_{H})}_{({\bf 1},{\bf 1},{\bf1})} + 
{({\delta^{c}_{H}}+
{\delta_{H}})}_{({\bf 6},{\bf 1},{\bf1})}
\nonumber\\
& &\mbox{}+{(e^{c}_{H}+{\nu^{c}_{H}}+
{d^{c}_{H}}+
{u^{c}_{H}})}_{({\bf {\overline 4}},{\bf 2},{\bf1})}+conj. reps.
\end{eqnarray}
 A $D$-flat v.e.v. $\langle {\nu^{c}_{H}}\rangle =
 \langle {\overline{\nu}}^{c}_{H} \rangle =V_{R}$ breaks 
$SU(4)\times SU(2)_{R}$ down to $SU(3)_{C}\times U(1)_{Y}$. The nine 
surviving states are the sextets ${({\delta^{c}_{H}} 
+{\delta_{H}})}_{({\bf 6},{\bf 1},{\bf1})}$, 
${({\overline{\delta}}^{c}_{H} +
{\overline{\delta}}_{H})}_{({\bf 6},{\bf 1},{\bf1})}$ and the singlets 
${(n_{H})}_{({\bf 1},{\bf 1},{\bf1})}$, 
${({\overline{n}}_{H})}_{({\bf 1},{\bf 1},{\bf1})}$, which do not 
participate in the symmetry breaking anyway, a pair of colour-triplets 
$d^{c}_{H}$, ${\overline d}^{c}_{H}$ and a combination of singlets 
${{\nu}^{c}_{H}}+{\overline{\nu}}^{c}_{H}$.

 The Higgs sector superpotential, suitable for the desired symmetry 
breaking pattern will be taken to be the cubic superpotential (invariant
under the discrete symmetry $H\to -H$)
\begin{equation}
W_{H}={\lambda}{\cal H}{H^2} +
{\overline{\lambda}}{\overline{\cal H}}{{\overline{H}}^2}
\end{equation}
which leads to the Higgs mass terms 
\begin{eqnarray}
{(W_{H})}_{mass}&=&{\lambda}{V_X}{\delta^{c}_{H}}\delta_{H}+
{\lambda}{V_{R}}{d^{c}_{H}}{\Delta_{H}}+
{\overline{\lambda}}{V_{X}}{\overline{\delta^{c}_{H}}}{\overline{\delta_{H}}}
+{\overline{\lambda}}{V_R}{\overline{d^c_H}}{\overline{\Delta_H}}
+...
\end{eqnarray}
Two pairs of coloured triplets, namely $\delta_H$, $\delta^c_H$ 
and ${\overline{\delta_H}}$, ${\overline{\delta^c_H}}$, obtain 
a mass of $O(M_X)$ and two pairs, namely $d^c_H$, $\Delta_H$ and 
${\overline{d^c_H}}$, ${\overline{\Delta_H}}$, obtain a mass of 
$O(M_R)$. Below $M_R$, apart from the four singlet states 
$n_H$, ${\overline{n_H}}$, ${N_H}+{\overline{N_H}}$ and 
${{\nu}^c_H}+{\overline{{\nu}^c_H}}$, the renormalizable 
superpotential interactions in $(3)$ leave the pair of coloured 
triplets ${\Delta}^c_H$, ${\overline{{\Delta}^c_H}}$ massless.
 Effective non-renormalizable terms could be added to $(3)$ which would 
generate masses for these states. These could arise due to the exchange 
of massive states present in all string constructions. Their exact form 
depends on the details of the specific string model. A necessary constraint 
on these terms is the requirement that they should not spoil $F$-flatness. 
 Note however that violations of $F$-flatness that lead to scalar masses 
of $O(TeV)$ can be tolerated since the supersymmetry breaking is of that 
order. This implies ${\langle F \rangle }/M\leq O(TeV)$ or 
${\langle F \rangle }\leq ({10}^{11}GeV)^2$. High order non-renormalizable 
terms of the form $({\cal H}{\overline{\cal H}})^{n}/{M}^{2n-3}$ lead to 
 $F\sim {({M_X}/M)^{2n-1}}{M^2}$ and satisfy this constraint if 
$({M_X}/M)^{2n-1}\leq {10}^{-14}$. For $M_X/M\sim 0.01$ 
this corresponds to $n\geq 4$. 
 Writing down the contents of such a term, gives 
\begin{equation}
{\Delta}W_H=({\cal H}{\overline{\cal H}})^{n}/{M^{2n-3}}=
{\frac{{(V_X)}^{2n-2}}{M^{2n-3}}}
({\Delta_H}{\overline{\Delta_H}}+{\Delta^c_H}{\overline{\Delta^c_H}}+\dots)
\end{equation}
This corresponds to an intermediate mass ${M_I}\sim {({M_X}/M)^{2n-2}}M$ 
for the triplets ${\Delta}^c_H$, ${\overline{\Delta^c_H}}$ and the 
singlet ${N_H}+{\overline{N_H}}$ . 
The generated intermediate mass is constrained to 
be $M_I\sim(M_X/M)^{2n-2}M\leq O(10^{-14})M^2/M_X$.
 For $M_X/M\sim 0.01$, this implies 
$M_I\leq O(10^6 GeV)$. 
For $M_X/M\sim 0.1$, $M_I\leq O(10^{5}GeV)$.
 Note also that, although the product in $(5)$ can 
arise either as ${(15)}_{ij}{({\overline{15})}}^{ij}$ or as 
${(15)}_{ij}{(15)}_{kl}{(15)}_{mn}{\epsilon}^{ijklmn}
{({\overline{15}})}^{pq}{({\overline{15}})}^{rs}{({\overline{15}})}^{tz}
{\epsilon}_{pqrstz}$ , only the first case gives rise to masses.
 It should be remarked however that the exact form of the non-renormalizable 
terms depends on the details of the, possibly, underlying string model and 
the above presented mechanism that supplies the remaining colour-triplets 
with an intermediate mass, although perfectly consistent in the framework 
of the GUT, serves as an existence proof of phenomenologically desirable 
scenaria that can be realized within the string model. 
 Summarizing, we see that introducing the Higgses
 $\cal H$, $\cal {\overline H}$ , $H$ , $\overline H$ we can achieve 
the successive breaking 
$SU(6)\times SU(2)_{L}\rightarrow SU(4)\times SU(2)_{R}\times SU(2)_{L}
\rightarrow SU(3)_{C}\times U(1)_{Y}\times SU(2)_{L}$ . Below $M_{X}$ 
the two pairs of  Higgses contribute to the spectrum of the model with 
two sextets $({\Delta_H}+{\Delta^c_H})_{({\overline{6}}, 1, 1)}$ and 
{\it {conjugate states}}, two 
tetraplets $({u^c_H}+{d^c_H}+{e^c_H}+{{\nu}^c_H})_{({\overline{4}}, 2, 1)}$
 and {\it {conjugate states}}, 
and the three singlets $n_H$, ${\overline{n_H}}$,
${N_H}+{\overline{N_H}}$. Below $M_R$ only the coloured triplets 
$\Delta^c_H$, ${\overline{\Delta^c_H}}$ survive, together with four 
singlets $n_H$, ${\overline{n_H}}$, ${N_H}+{\overline{N_H}}$, 
${{\nu}^c_H}+{\overline{{\nu}^c_H}}$. The coloured triplets obtain 
a mass at the lower scale $M_I$. 

 Quarks and leptons can be accommodated in the $({\bf {\overline {15}}},{\bf 1})
+({\bf 6},{\bf 2})$ representation in three family replicas
\begin{eqnarray}
\phi({\bf {\overline {15}}},{\bf 1})&=&
N_{({\bf 1},{\bf 1},\bf{1})}+
{({\delta^{c}}+{\delta})}_{({\bf  6},{\bf 1},\bf{1})}
+{({e^{c}}+{\nu^{c}}+{d^{c}}+{u^{c}})}_{({\bf {\overline 4}},{\bf 2},\bf{1})}
\\
\psi({\bf 6},{\bf 2})&=&
{(h+{h}^{c})}_{({\bf 1},{\bf 2},{\bf 2})} +{(q+l)}_{({\bf 4},{\bf 1},{\bf 2})}
\end{eqnarray}
 Notice that together with quarks and leptons we have pairs of isodoublets, 
suitable as 
electroweak Higgses $h$ , $h^{c}$ and coloured triplets $\delta$ , 
$\delta^{c}$ in three family replicas too. 
We next introduce the matter self coupling
\begin{eqnarray}
W_{M}&=&{Y}_{ijk}{\phi_i}{\psi_j}{\psi_k}\nonumber\\
&=&{\tilde{Y}}_{ijk}({q_j}{d^{c}_{i}}h_k+{q_j}{u^{c}_{i}}{h^{c}_{k}}
+{l_j}{e^{c}_{i}}{h_k}+{l_j}{{\nu}^{c}_{i}}{h^{c}_{k}}+
{\frac{1}{2}}{{\delta}_{i}}{q_j}{q_k}
+{{\delta}^{c}_{i}}{q_j}{l_k}+
{N_i}{h_j}{h^{c}_{k}})
\end{eqnarray}
where ${\tilde{Y}}_{ijk}={Y_{ijk}}+{Y_{ikj}}$. 
Quark and lepton masses arise exclusively from the single 
Yukawa coupling ${\tilde{Y}}_{ijk}$. 
 The fifth and sixth term in $(8)$, together with 
a ${\delta}{\delta^c}$ mass term for the extra colour-triplets, give rise to 
the $D=5$ operators $qqql$ that violate baryon number and can induce proton 
decay \cite{PRDEC}. These operators are controllable if 
the ${\delta_i}{\delta^c_j}$ mass is of $O(M_X)$. This can be achieved 
if we introduce the couplings 
\begin{equation}
{\Delta W}_{M}={\frac{1}{2}}{\lambda}_{ij}{\cal H}{{\phi}_{i}}{{\phi}_{j}}
+{\frac{1}{2}}{\lambda}^{\prime}_{ij}{\cal H}{\psi_i}{\psi_j}
\end{equation}
which give the terms 
$$
{\lambda_{ij}}({N_H}{\delta_i}{\delta^c_j}+
{e^{c}_{j}}{u^{c}_{j}}{\Delta_H}
+{{\nu}^{c}_{i}}{d^{c}_{j}}{\Delta_H}+
{u^{c}_{i}}{d^{c}_{j}}{\Delta^{c}_{H}}+...)
$$
\begin{equation}
+{\lambda^{\prime}_{ij}}({N_H}{h_i}{h^c_j}+
{q_i}{q_j}{\Delta_H}+{q_i}{l_j}{\Delta^c_H})
\end{equation} 
 Masses of $O({\lambda}V_X)$ and $O({\lambda^{\prime}}V_X)$ correspondingly 
can now arise for the coloured triplets and isodoublets. Note 
that both $\lambda_{ij}$ and $\lambda^{\prime}_{ij}$ couplings are symmetric and 
$\lambda_{ij}'$ must have at least one zero eigenvalue.
 Baryon number violating operators of $D=5$ of the type 
${e^c}{u^c}{u^c}{d^c}$, ${\nu^c}{u^c}{d^c}{d^c}$ and $qqql$ would also appear 
if a mass term $\Delta_H$-$\Delta^c_H$ were present. This is not the 
case however since $\Delta^c_H$ mixes only with ${\overline{\Delta^c_H}}$ 
which does not mix with matter fields. 
Nevertheless, $D=6$ operators 
 of the form 
$\lambda_{ij}\lambda_{km}'
\left(u^c_i d^c_j q^\dagger_k l^\dagger_m\right)$,
generated by the exchange of a scalar triplet
${\Delta_H}$ or
${\Delta^c_H}$, are still possible and can be dangerous in the case of 
an intermediate
triplet mass.  The presence of these operators depends on the  structure of
the coupling matrices  
$\lambda_{ij}\lambda_{km}'$. For instance, in the case of diagonal couplings, 
a simple
condition like $\lambda_{1j}=0,j=1,2$ would be sufficient to render 
these operators
harmless without affecting the massiveness of $\delta_i,\delta^c_i$. 
In a general  situation this problem can be 
evaded by
requiring  a  more complicated  texture
structure  for $\lambda_{ij}$ and/or $\lambda_{km}'$. Another more drastic 
solution is to set 
$\lambda_{km}'=0$. In this case no such operators exist but all three pairs
of  isodoublets are left massless. This is not a big problem since their 
masses can be
generated by non-renormalizable terms. For example the term
$\frac{\lambda_{ij}''}{M^3}H^4\psi_i\psi_j$ leads to a 
${\lambda_{ij}''}\frac{<\nu^c_H>^4}{M^3} h_i h^c_j$ isodoublet mass matrix without introducing
any unwanted couplings.

Expressing the Yukawa interactions in terms of the Higgs mass eigenstates, 
we obtain
\begin{equation}
{Y_{ij}^{a}}({q_i}{d^c_j}{\tilde{h}}_a+{l_i}{e^c_j}{\tilde{h}}_{a}+
{q_i}{u^c_j}{\tilde{h^c}_a}+{l_i}{\nu^c_j}{\tilde{h^c}_a}), a=1,2,3
\end{equation}
 It is clear that in the case of only one pair of massless Higgses, since 
 there is a single, generation dependent, Yukawa coupling, 
we have the fermion mass relations ${m_b}={m_t}cot{\beta}=m_{\tau}$ 
which imply the prediction of $m_t$ \cite{LLL} as well as strong 
constraints on $tan{\beta}$ ($tan{\beta}\geq 40$) and the 
susy-breaking parameters \cite{WWW}. In the case that a second pair 
of isodoublets is massless and obtain v.e.v.'s, 
there is no simple mass relation unless there is v.e.v. alignment.
 Depending on the details of the specific string model or possibly 
existing family symmetries, the Yukawa couplings for the light 
generations could arise as effective non-renormalizable terms with 
a suppressing factor $(\langle S \rangle /M)^{n}$ involving 
the v.e.v. of a singlet field $S$ \cite{JRKT}.
 Nevertheless, an alternative to such a scenario would be 
to keep more doublets light and impose a hierarchy on the 
available many v.e.v's.
 Non-renormalizable terms can also  
generate the required large Majorana mass for the right-handed 
neutrino. Introducing 
\begin{equation}{Y''_{ij}}{\phi_i}{\phi_j}{\overline{H}}^{2}/M\end{equation}
we are only led to 
\begin{equation}Y''_{ij}{\nu^{c}_{i}}{\nu^{c}_{j}}
{\langle {\overline{\nu}}^{c}_{H}\rangle }^{2}/M=
{\frac {{Y''_{ij}}V^{2}_{R}}{M}}
{{\nu}^{c}_{i}}{{\nu}^{c}_{j}}\end{equation}
Depending on the effective non-renormalizable coupling $Y''_{ij}$ 
this term can lead to acceptable neutrino masses through a see-saw 
mechanism.

\medskip
\begin{it}b) An $SU(6)\times SU(2)_{R}$ model.\end{it}

 The first breaking $SU(6)\times SU(2)_{R}\rightarrow 
SU(4)\times SU(2)_{L}\times SU(2)_{R}$ occurs 
exactly as in the previous model using a pair of Higgses 
$({\bf 15},{\bf 1})+({\bf {\overline{15}}},{\bf 1})$ . Nevertheless, 
due to the different position of $SU(2)_{L}$ their content looks 
different in terms of standard model quantum numbers. 
Again we introduce
\begin{equation}
{\cal H}({\bf 15},{\bf 1})+
{\cal {\overline H}}({\bf {\overline {15}}},{\bf 1})=
{(N_{H})}_{({\bf 1},{\bf 1},{\bf1})}+{({\Delta_H}+
{\Delta^{c}_{H}})}_{({\bf 6},{\bf 1},{\bf1})}
+{({L_H}+{Q_H})}_{({\bf 4},{\bf 2},{\bf1})}+conj. reps.
\end{equation}
  A $D$-flat v.e.v. $\langle {N_H} \rangle =
\langle {\overline N}_{H} \rangle =V_X$ breaks the gauge group 
 to $SU(4)\times SU(2)_{L}\times SU(2)_{R}$. 
The coloured triplets $\Delta_H$ , $\Delta^{c}_{H}$ , 
$\overline {\Delta}^{c}_{H}$ , $\overline{\Delta}_H$
and a combination of singlets, namely $N_{H}+{\overline N}_{H}$, 
survive the Higgs phenomena. 

The second stage of symmetry 
breaking down to the Standard Model group is achieved 
introducing 
\begin{equation}H({\bf  6},{\bf 2})+
{\overline H}({\bf 6},{\bf 2})=
{({\eta_H}+{\eta^{c}_{H}})}_{({\bf 1},{\bf 2},{\bf 2})}
+{({d^{c}_{H}}+{u^{c}_{H}}+{e^{c}_{H}}
+{\nu^{c}_{H}})}_{({\bf {\overline 4}},{\bf 1},{\bf 2})}
+conj. reps.
\end{equation}
 A $D$-flat v.e.v. $\langle {\nu^{c}_{H}}\rangle =
\langle {\overline{\nu}}^{c}_{H} \rangle =V_R $ breaks the 
gauge group down to $SU(3)_{C}\times SU(2)_{L}\times U(1)_{Y}$.
 In addition to $\eta_H$ , $\eta^{c}_{H}$ , ${\overline{\eta}}_H$ , 
${\overline{\eta}}^{c}_{H}$ which are not affected 
by the Higgs phenomena, $d^{c}_{H}$ , ${\overline{d}}^{c}_{H}$ and 
the singlet ${\nu^{c}_{H}}+{\overline{\nu}}^{c}_{H}$ survive.

 The Higgs sector superpotential will be taken to be again a cubic 
one of the form
\begin{equation}
W_H={\lambda}{\cal H}H^2+
{\overline{\lambda}}{\overline{\cal H}}{\overline{H}}^2
\end{equation}
which leads to the Higgs mass terms 
\begin{equation}
(W_H)_{mass}={\lambda}{V_X}{\eta_H}{\eta^c_H}+
{\lambda}{V_R}{\Delta_H}{d^c_H}+conj. reps.
\end{equation}
Two pairs of Higgs isodoublets become massive at $M_X$ and 
two pairs of coloured triplets, namely $\Delta_H$, $d^c_H$ and 
${\overline{\Delta_H}}$, ${\overline{d^c_H}}$ become massive at $M_R$.
 Below $M_R$ we are left with a massless pair of colour-triplets 
$\Delta^c_H$, $\overline{\Delta^c_H}$. Following the philosophy of the 
previous model we can expect that non-renormalizable 
terms  $({\cal H}{\overline{\cal H}})^{n}/M^{2n-3}$ will give 
rise to a mass of intermediate value
 $M_I\sim {10}^{6} GeV$ or smaller for the 
pair $\Delta^c_H$, $\overline{\Delta^c_H}$ as well as for the 
singlet $N_H+{\overline{N_H}}$.

 The matter fields are introduced as 
\begin{eqnarray}{\phi}({\bf 15},{\bf 1})&=&N_{({\bf 1},{\bf 1},{\bf1})}+
{({\delta}+{\delta^{c}})}_{({\bf 6},{\bf 1},{\bf1})}+
{(l+q)}_{({\bf 4},{\bf 2},{\bf1})}\\
{\psi}({\bf \overline{6}},{\bf 2})&=&
{(h+{h}^{c})}_{({\bf 1},{\bf 2},{\bf 2})}
+(d^{c}+u^{c}+e^{c}+
{\nu}^{c})_{({\bf {\overline 4}},
{\bf 1},{\bf 2})}\end{eqnarray}
in three family replicas. The Yukawa terms are obtained from
\begin{eqnarray}
W_M&=&{Y_{ijk}}{\phi_i}{\psi_j}{\psi_k}\\
&=&\mbox{}{\tilde{Y}_{ijk}}({l_i}{e^c_j}{h_k}
+{l_i}{\nu^c_j}{h^c_k}
+{q_i}{d^c_j}{h_k}+{q_i}{u^c_j}{h^c_k}
+{N_i}{h_j}{h^c_k}+{\delta_i}{d^c_j}{\nu^c_k}+
{\delta_i}{u^c_j}{e^c_k}+{\delta^c_i}{d^c_j}{u^c_k})
\end{eqnarray}
 Let us now introduce the Higgs-matter interactions
\begin{equation}
{\Delta}W_M={\lambda_{ij}}{\cal H}{\phi_i}{\phi_j}
+{\lambda^{\prime}_{ij}}{\cal H}{\psi_i}{\psi_j}
\end{equation}
$$={\lambda_{ij}}({N_H}{\delta_i}{\delta^c_j}+
{\Delta_H}{q_i}{q_j})+{\lambda^{\prime}_{ij}}({N_H}{h_i}{h^c_j}
+{\Delta_H}{d^c_i}{\nu^c_j}+
{\Delta_H}{u^c_i}{e^c_j}+{\Delta^c_H}{d^c_i}{u^c_j})$$
 Note that the couplings 
$\lambda_{ij}$, $\lambda^{\prime}_{ij}$ are symmetric. 
These couplings are sufficient to render all triplets 
$\delta_i$, $\delta^c_i$ massive with a mass of $O(M_X)$.
 Of course $\lambda^{\prime}_{ij}$ has to be restricted in 
family space in order to obtain the 
desired number of massless pairs of electroweak Higgses. 
The $D=5$ operators that violate baryon number and can arise 
from $W_M$ are going to be sufficiently suppressed if 
$\delta_i$, $\delta^c_i$ have a mass of $O(M_X)$. On the 
other hand, since $\Delta_H$ and $\Delta^c_H$ do not mix, 
there is no danger of any $D=5$ operators arising from 
$\Delta W_M$. Note that $\Delta_H$ obtains its mass from 
$d^c_H$ which does not mix with matter. D=6 operators
involving the exchange of a $\Delta_H$ scalar are also possible
and could be dangerous for low $M_R$. The related discussion in
the previous model applies also here. If the condition
$\lambda_{ij}'=0$ is imposed in order to circumvent  this problem,
the  non-renormalizable term
$\frac{\lambda_{ij}''}{M^3}H^4\psi_i\psi_j$ can be invoked to generate 
$\frac{<\nu^c_H>^4}{M^3}$ masses for the unwanted doublets.

 Finally the Yukawa interactions are identical to those 
of the previous model and all points concerning Yukawa coupling 
unification and mass relations are the same. Again, in order to 
obtain acceptable neutrino masses, the non-renormalizable 
interactions 
$${Y^{{\prime}{\prime}}_{ij}}{\psi_i}{\psi_j}{\overline{H}}
{\overline{H}}/M$$
have to be invoked. This term generates a Majorana mass matrix for 
the right-handed neutrinos of 
order ${Y^{{\prime}{\prime}}_{ij}}{V^2_R}/M$ .
\bigskip

\begin{bf}3. Renormalization group analysis.\end{bf}

 As follows from the previous analysis the $SU(2)_R$ scale $M_R$ is
a free parameter while the triplet mass scale $M_I$
depends on the details of the model and in particular of the
non-renormalizable contributions in the superpotential. 
In addition the $SU(6)\times SU(2)_{L}$ model, 
denoted as  model (I) from now 
on, and the $SU(4)\times SU(2)_{R}$ model, denoted as  model (II),
 are indistinguishable bellow $M_{R}$ where
 we have an $SU(3)_{C}\times 
SU(2)_{L}\times U(1)_{Y}$ theory with an extra charge-1/3 colour-triplet pair 
of intermediate mass $M_{I}$.  In what follows we shall assume both
$M_R$ and $M_I$ as free parameters and derive the constraints imposed 
on them when demanding $SU(6)$ and $SU(2)$ coupling unification
at the string scale $M_S = 5.27\times g_S\times 10^{17} GeV$. For simplicity we
consider here only the case of one or two ($N_H=1,2$) massless isodoublet pairs
below $M_R$ and leave the more complicated analysis of extra intermediate
mass  doublets for a future publication \cite{PREP}.
 Integrating the renormalization 
group equations from $M_{Z}$ up to $M_{R}$, we obtain
\begin{equation}{\alpha}^{-1}_{i}(M_{R})={\alpha}^{-1}_{i}+
{\frac {b_{i}}{2{\pi}}}{\ln ({\frac {M_{Z}}{M_{R}}})}+
{\frac {{\tilde{b}}_{i}}{2{\pi}}}{\ln ({\frac {M_{I}}{M_{R}}})}\end{equation}
 Where $i=1,2,3$ and $\alpha_1$ , $\alpha_2$ , $\alpha_3$ 
stand for the values of the three gauge couplings at $M_Z$.
 The RG coefficients take up the values $b_{3}=-3$ ,
 $b_{2}=N_H$ , $b_{1}=6+{\frac{3N_H}{5}}$ , 
${\tilde{b}}_{3}=1$ , ${\tilde{b}}_{2}=0$ ,${\tilde b}_1=\frac{2}{5}$
 At an energy scale ${\mu}\ge M_R$ we have an  
$SU(4)\times SU(2)_{R}\times SU(2)_{L}$ theory
 in both cases (I) and (II). Note that the particle content 
for (I) and (II) is the same. The three 
gauge couplings are given by 
\begin{equation}
{\alpha}^{-1}_{i}({\mu})={\alpha}^{-1}_{i}(M_{R})+
{\frac {b_{i}}{2{\pi}}}{\ln ({\frac {M_{R}}{\mu}})}
\end{equation}
Where $i=4,\, 2R,\, 2L$ . The coefficients take up the values 
$b_{4}=-2$ , $b_{2L}=N_H$ , $b_{2R}=4+N_H$.
 The matching conditions at $M_R$ are 
\begin{eqnarray}{\alpha}_{4}(M_{R})&=&{\alpha}_{3}(M_{R})\\
{\alpha}^{-1}_{2R}(M_{R})&=&
{\frac {5}{3}}{\alpha}^{-1}_{1}(M_{R})
-{\frac {2}{3}}{\alpha}^{-1}_{3}(M_{R})\\
{\alpha}_{2L}(M_{R})&=&{\alpha}_{2}(M_{R})\end{eqnarray}
 As $\mu$ moves away from $M_{R}$ two possibilities arise. 
If ${\alpha_{2R}}$ meets first with $\alpha_4$, we have model (I). 
In the opposite case of $\alpha_{2L}$ meeting first with $\alpha_4$, we 
have model (II). From the point of view of the
renormalization group analysis,  the low energy data as well as 
the free parameters 
of the theory,  $M_{R}$ and $M_{I}$, 
determine which case is realized. 

 Let us first study the case of model (I) which is defined by
\begin{equation}
\alpha_6(M_X) = {\alpha}_{2R}(M_{X})=
{\alpha}_{4}(M_{X})\geq {\alpha}_{2L}(M_{X})
\label{edi}
\end{equation}
$M_{X}$ stands for the $SU(6)\times SU(2)$ breaking scale. Manipulating all 
previously displayed equations, we obtain 
\begin{equation}
{\frac {1}{2\pi}}{\ln ({\frac {M_{X}}{M_{Z}}})}=
{\frac {5}{3(6+N_H)}}({\alpha}^{-1}_{1}-{\alpha}^{-1}_{3})-
{\frac {4}{{\pi}(6+N_H)}}\ln ({\frac {M_{R}}{M_{Z}})}-
{\frac {1}{2{\pi}(6+N_H)}}
{\ln ({\frac {M_{I}}{M_{Z}}})}
\label{exi}
\end{equation}
The inequality in (\ref{edi}) defining model (I) is equivalent to 
\begin{equation}
F={\frac {5}{3}}(2+N_H){\alpha}^{-1}_{1}+{\frac {2}{3}}(4-N_H){\alpha}^{-1}_{3}
-(6+N_H){\alpha}^{-1}_{2}
+{\frac {2}{\pi}}{\ln ({\frac {M_{I}}{M_{Z}}})}
-{\frac {4}{\pi}}(2+N_H){\ln ({\frac {M_{R}}{M_{Z}}})}\leq 0
\end{equation}

\begin{centering}
\begin{figure}[ht]
\epsfxsize=14.5cm
\epsfbox[0 240 560 590]{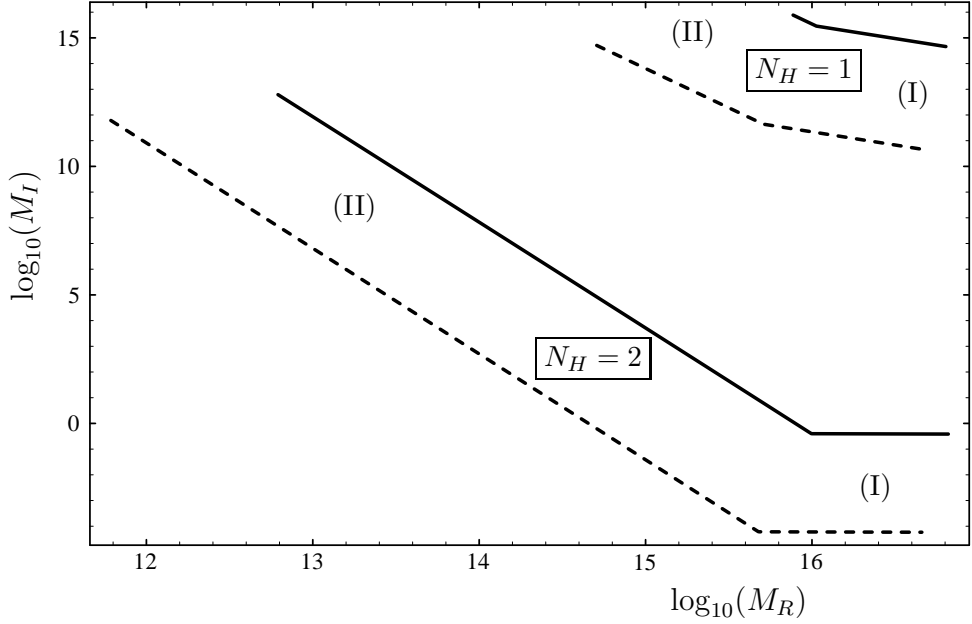}
\caption{ The intermediate scale $\log_{10}(M_I)$ in the case 
$N_H=1,2$ as function of 
$\log_{10}(M_R)$, for
 $\alpha_3=.11$ (dashed line) and $\alpha_3=.13$ (continuous line).}
\vskip -1.6cm \hskip 10cm $\log_{10}(M_R)$
\vskip -6cm  \hskip 1.2cm\rotate[l]{ $\log_{10}(M_I)$}
\vskip -4.1 cm \hskip 10 cm {\small(II)}
\vskip  -.01cm \hskip11cm \fbox{{\small$N_H=1$}}
\vskip -.2cm \hskip 13 cm {\small(I)}
\vskip   1cm \hskip 5.5 cm {\small(II)}
\vskip  1.5cm \hskip8.2cm {\fbox{\small$N_H=2$}}
\vskip  1.2cm \hskip 12.5 cm {\small(I)}
\vskip 2.5 cm
\end{figure}
\end{centering}

 At a scale ${\mu}\ge M_{X}$ the $SU(6)\times SU(2)_{L}$ gauge 
couplings are 
\begin{equation}
{\alpha}^{-1}_{i}({\mu})={\alpha}^{-1}_{i}(M_{X})+
{\frac {b_{i}}{2{\pi}}}{\ln ({\frac {M_{X}}{\mu}})}
\end{equation}
Where $i=6,\, 2L$ .  The coefficients take up the values
 $b_{6}=-1$ , $b_{2L}=3$ . The fact that $b_6<0$ and $b_{2L}>0$
 together with (\ref{edi}) imply that there exists always a scale ${\mu}=M_S>M_X$
 at which the $SU(6)$ and $SU(2)$ couplings become equal 
\begin{equation}
{\alpha}_{6}(M_{S})=
{\alpha}_{2L}(M_{S})\equiv {\alpha}_{S}\end{equation}
Solving for it, we obtain 
\begin{equation}
{\frac {1}{2{\pi}}}{\ln ({\frac {M_{S}}{M_{Z}}})}=
{\frac {1}{4}}({\alpha}^{-1}_{2}-{\alpha}^{-1}_{3})
-{\frac {1}{8\pi}}{\ln ({\frac {M_I}{M_Z}})}+
{\frac{(2-N_H)}{8\pi}}{\ln({M_X}/{M_Z})}
\end{equation}
while the value of the unified gauge coupling  is given by
\begin{equation}
\alpha^{-1}_{S}=
\alpha^{-1}_{3}+{\frac {1}{2{\pi}}}{\ln {\frac {M_S}{M_Z}}}
+{\frac {1}{2\pi}}{\ln {\frac {M_X}{M_Z}}}+
{\frac {1}{2\pi}}{\ln {\frac {M_I}{M_Z}}}
\end{equation}

 The case of the model (II) is obtained  for
\begin{equation}
\alpha_6={\alpha}_{2L}(M_{X})={\alpha}_{4}(M_{X})\geq {\alpha}_{2R}(M_{X})
\end{equation}
and the inequality (\ref{edi}) amounts to $F\ge0$.
The GUT scale is given by 
\begin{equation}
{\frac {1}{2{\pi}}}{\ln ({\frac {M_{X}}{M_{Z}}})}=
{\frac {1}{(N_H+2)}}({\alpha}^{-1}_{2}-{\alpha}^{-1}_{3})
-{\frac {1}{2{\pi}(N_H+2)}}{\ln ({\frac {M_I}{M_Z}})}
\label{exii}
\end{equation}
 The $SU(6)\times SU(2)_{R}$ gauge couplings at a scale ${\mu}\ge M_X$ 
are given by the corresponding expressions for model (I) with the 
replacements $b_{6}\rightarrow b'_{6}=-3$ and $b'_{2L}\rightarrow 
b'_{2R}=9$. These 
couplings always meet at a scale $M_{S}$ defined by
\begin{equation}
{\alpha}_{6}(M_{S})=
{\alpha}_{2R}(M_{S})\equiv {\alpha}_{S}\end{equation}
The scale $M_{S}$ is obtained to be 

\begin{equation}
{\frac {1}{2{\pi}}}{\ln ({\frac {M_{S}}{M_{Z}}})}=
{\frac{5}{36}}({\alpha^{-1}_{1}}-{\alpha^{-1}_{3}})-
{\frac{1}{24{\pi}}}{\ln ({M_I}/{M_Z})}+
{\frac {(6-N_H)}{{24{\pi}}}}{\ln ({M_X}/{M_Z})}-
{\frac {1}{3{\pi}}}{\ln ({M_R}/{M_Z})}
\end{equation}
while the formula for the unified gauge coupling takes the form
\begin{equation}
\alpha^{-1}_{S}=
\alpha^{-1}_{3}+{\frac {3}{2\pi}}{\ln {\frac {M_S}{M_Z}}}
+{\frac {1}{2\pi}}{\ln {\frac {M_I}{M_Z}}}-
{\frac {1}{2\pi}}{\ln {\frac {M_X}{M_Z}}}
\label{eaii}
\end{equation}
%
%
Given $M_I$ and $M_R$ and the values of the low energy parameters
we can now check whether $F<0$ or $F>0$. 
In the former case we  are in Model (I) and
we can use (29), (33), (34) in order to solve for 
$M_X$ and $M_S$ as well as $\alpha_S$. In  the latter
we are in model (II) and we must use (37), (39), (40)
instead.
If we further assume that the unification scale $M_S$ is the 
{\em string scale} 
$M_S = 5.27\times g_S\times 10^{17} GeV$ 
we can
express $M_I$, and subsequently $M_X$ and $\alpha_S$, as a function of $M_R$.

 In figures 1,2 and 3 we give plots of $M_I$ and $M_X$ as 
functions of  $\log(M_R)$ for $M_S = 5.27\times g_S\times 10^{17} GeV$ 
in the cases $N_H=1,2$ for both models (I) and (II). 
As indicative input values 
we have taken those of \cite{HHH}, 
while for $\alpha_3(M_Z)$ we have utilized 
the range 0.11-0.13.
 We shall not worry about the  uncertainty in these values 
since we intend here to investigate the
 generic dependence  on the  parameter 
$M_R$  instead of obtaining a precise determination.
The allowed region is the one between the dashed and 
the full line.  
As it can be seen in the figures there is a wide range of $M_R$ and $M_I$
values for which both versions of the model can be realized. 
Namely, for $N_H=1$ model (I) requires\footnote{We have taken into account proton
decay constraints on $M_X$.}
$10^{15.8} GeV< M_R < 10^{16.3}GeV$
and model (II) 
$10^{14.6}GeV < M_R < 10^{16.1}GeV$
while 
$10^{11}GeV < M_I < 10^{15}GeV$.
For $N_H =2 $, model (I)  demands
$10^{15.4}GeV < M_R < 10^{16.4}GeV$ and $M_I\sim{\cal O} (GeV)$ 
and model (II) 
$10^{14}GeV < M_R < 10^{15.8}GeV$ with  
$10^{3}GeV < M_I < 10^{8}GeV$.
From fig. 1 it can be seen that the intermediate scale $M_I$
in the case of model (I) is either very small ($N_H=2$) or very large ($N_H=1$).
The previously proposed mechanism for generation of  $M_I$ through
the specific non-renormalizable terms (5), can be realized for $M_R\sim 10^{14}-
10^{15} GeV$, $M_I\sim 10^{6} GeV$ in the case of model (II).
Note however that we have left for future study \cite{PREP} the more 
complicated case
of an intermediate scale mass for the second pair of Higgs isodoublets 
which could lead to acceptable $M_I$ within the range of the proposed mechanism 
for both models.

\begin{centering}
\begin{figure}[ht]
\epsfxsize=14.5cm
\epsfbox[0 240 560 590]{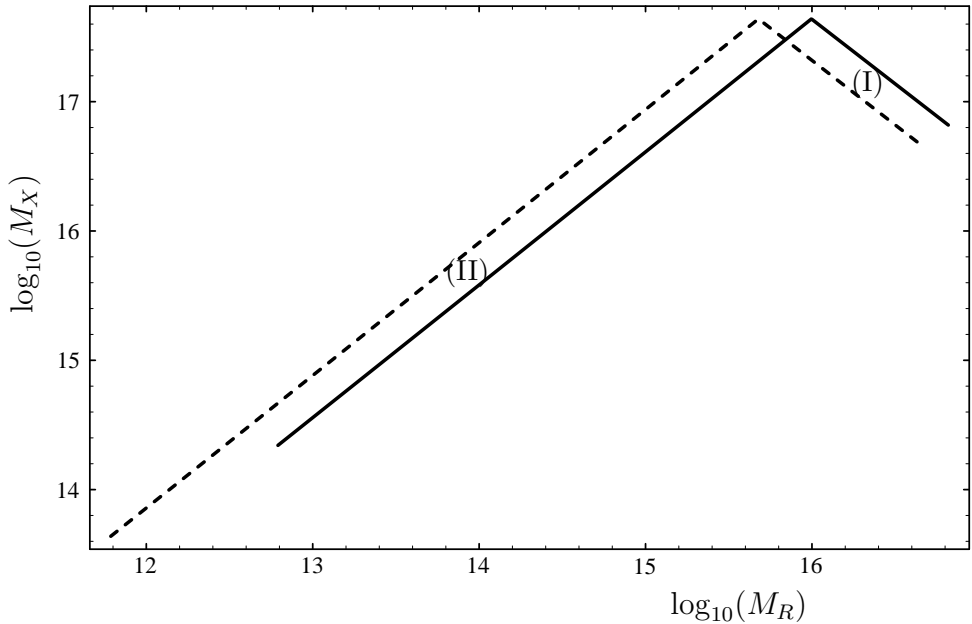}
\caption{ The GUT scale $\log_{10}(M_X)$ in the case $N_H=2$ as function of 
$\log_{10}(M_R)$,
 for $\alpha_3=.11$ (dashed line) 
and $\alpha_3=.13$ (continuous line).}
\vskip -1.6cm \hskip 10cm $\log_{10}(M_R) $
\vskip -6cm  \hskip 1.2cm\rotate[l]{ $\log_{10}(M_X)$}
\vskip -3.6 cm \hskip 12.4 cm {\small (I)}
\vskip 2cm \hskip 7 cm {\small(II)}
\vskip 6 cm
\end{figure}
\end{centering}

\begin{centering}
\begin{figure}[ht]
\epsfxsize=14.5cm
\epsfbox[0 240 560 590]{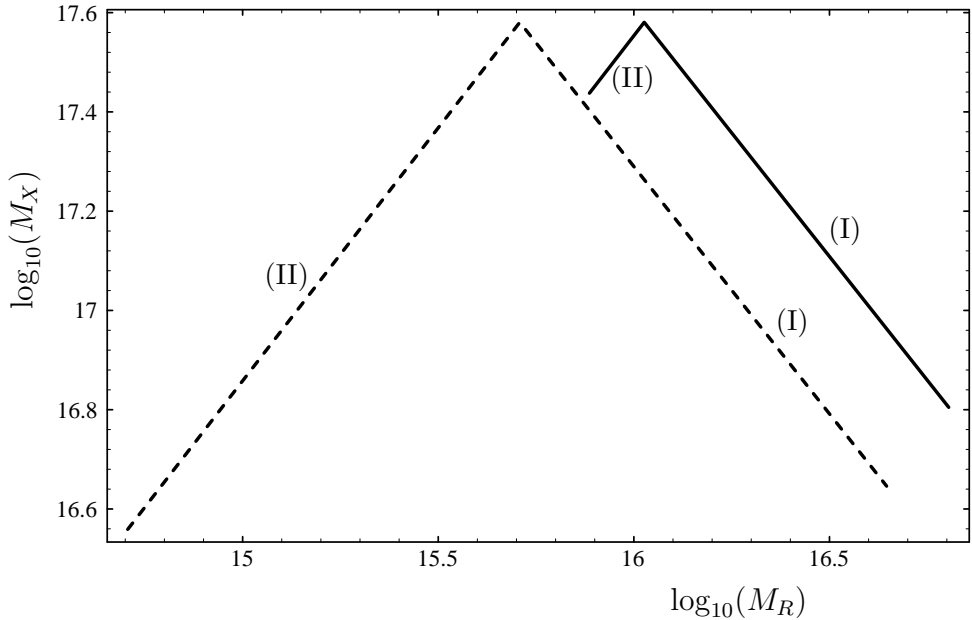}
\caption{ The GUT scale $\log_{10}(M_X)$ in the case $N_H=1$ as function of 
$\log_{10}(M_R)$, 
for $\alpha_3=.11$ (dashed line) and $\alpha_3=.13$ (continuous line).}
\vskip -1.6cm \hskip 10cm $\log_{10}(M_R)$
\vskip -6cm  \hskip 1.2cm\rotate[l]{ $\log_{10}(M_X)$}
\vskip -3.6cm \hskip 9.2 cm {\small(II)}
\vskip  1.5cm \hskip 12.1  cm {\small(I)}
\vskip .1cm \hskip 4.6 cm {\small (II)}
\vskip .1cm \hskip 11.4 cm {\small(I)}
\vskip 5 cm
\end{figure}
\end{centering}
\bigskip
\begin{bf} 4. Brief discussion and conclusions.\end{bf}

 We have constructed and analyzed $SU(6)\times SU(2)$ GUT. Depending
on the identification of the electroweak $SU(2)_L$ gauge group with
the GUT $SU(2)$ factor or with the $SU(2)\subset SU(6)$ two possible
models, $SU(6)\times SU(2)_L$ and $SU(6)\times SU(2)_R$, arise.
Both models can accommodate the MSSM particles, as well as the
gauge symmetry breaking Higgses, in representations allowed by the
$k=1$ superstring embedding. The models possess 
several characteristic features of their own. From a group 
theoretic point of view they are unique since only 
$({\bf 15},{\bf 1})$ and $({\bf {\overline{6}}},{\bf 2})$ 
representations are sufficient to accommodate matter as well as 
all the Higgses required for symmetry breaking. In contrast, 
flipped $SU(5)$ or minimal $SU(5)$ need $({\bf 10},1)$ , 
$({\bf {\overline{5}}},-3)$ and $({\bf 5},-2)$ or ${\bf 10}$ , 
${\bf 5}$ and ${\bf 24}$ . The models are characterized by an 
intermediate, nevertheless high, scale $M_R$. The gauge couplings 
above the {\it {unification scale}} $M_X$ always intersect at a 
scale $M_S$ which for a range of values of $M_R$ can play the role 
of the {\it {string unification scale}}. Thus, for these models 
there is no unification mismatch problem. The particle content 
of the resulting $SU(3)_{C}\times SU(2)_{L}\times U(1)_{Y}$ 
theory bellow $M_R$ is that of MSSM with an extra charge-1/3 
colour-triplet pair of mass $M_I$ and, possibly,
an extra pair of electroweak isodoublets. Remarkably
the intermediate scale $M_I$ can be relatively small as
it can be explicitly seen in the framework of the proposed mechanism
for the extra triplet mass generation.

The question of the generation of the scale $M_R$ and $M_X$, 
 a general question  addressing all scales that arise from flat directions,
could in principle be studied in the framework of the softly broken theory.
There, the supersymmetry breaking in conjunction with                         
non-renormalizable terms
can stabilize the related vevs and give rise to intermediate scales.
However, this is not an easy problem  since the susy breaking parameters run
appreciably with the renormalization scale. This study should            
necessarily involve the renormalization group equations for the soft
parameters \cite{GMR} and it is beyond the scope of this article.

 An interesting property of these models, inherited from $E(6)$, 
and shared by $SO(10)$, is the fact that they are characterized by
 {\it {Yukawa coupling unification}}, since all fermion masses 
result from a common generation-dependent Yukawa coupling. Note however 
that only in the case of one pair of electroweak 
v.e.v.s we obtain 
simple fermion mass relations. 
On the other hand, the Higgs isodoublets and 
the extra pair of right-handed 
d-quarks contained in the matter representations can receive masses 
separately and can be split without the need for any numerical 
fine tuning. In order to maintain one or more massless isodoublet 
pairs in a truly natural sense a family-dependent symmetry would be 
required. Nevertheless, if, as it happens in specific string constructions, 
the Yukawa coupling family hierarchy comes about as a hierarchy in the 
order of non-renormalizable terms, such a family symmetry could arise 
in the form of a selection rule \cite{JRKT}. 

 Let us now say a few things about the features of possible
 embedding of these models in  $k=1$ superstring constructions.
Assuming a gauge group $G=\prod_{i\in I}G_{i}$ , where 
the group factors are realized at the level $k_{i}$ of the 
{\it Affine world sheet algebra} , the conformal weight of 
a representation $(r_{1},...r_{n})$ of $G$ is given by the 
formula $h_{KM}=\sum_{i=1}^{n}{\frac {dim(G_{i})}{dim(r_{i})}}
{\frac {T(r_{i})}{(k_{i}+T(A_{i}))}}$ where $T(r_{i})$ is the 
index of the representation $r_{i}$ normalized as $T={\frac {1}{2}}$ 
for the vector of $SU(n)$. $A_{i}$ is the adjoint of $G_{i}$. 
Applying this formula for $G=SU(6)\times SU(2)$ and $k_{i}=1$, 
we obtain that the conformal weights of the 
representations $({\bf 6},{\bf 1})$ , $({\bf 6},{\bf 2})$ , 
$({\bf 15},{\bf 1})$ , $({\bf 15},{\bf 2})$ , $({\bf 20},{\bf 1})$ , 
$({\bf 20},{\bf 2})$ are 5/12, 2/3, 2/3, 11/12, 3/4 and 1 
correspondingly and thus all these states could be generically 
massless in a $k=1$ construction.   Some of these representations e.g
$({\bf 6},{\bf 1})$ will lead to exotic fractional charge states.
 In a fully realistic superstring derived model the appearing exotic 
representations  should circumvent conflicts with phenomenology either 
due to supermassiveness 
or due to the confining properties of the hidden sector gauge group.
For the moment  no ``realistic" $k=1$  $SU(6)\times SU(2)$ superstring model 
has been constructed
\footnote{For some toy models obtained in the context 
 $Z_6-II$ Orbifold see \cite{OOO}.}.
 However,
the GUT analysis provides encouraging results in order to proceed towards this
direction.

\centerline{\bf Acknowledgements}
We acknowledge financial support from the research program 
$\Pi{\rm ENE}\Delta$-95
of the Greek Ministry
of Science and Technology. We would like to thank C. Panagiotakopoulos for
discussions.

\end{document}